\documentclass[aps,prl,twocolumn,showpacs,superscriptaddress,notitlepage,longbibliography]{revtex4-2}
\usepackage{graphicx}
\usepackage{subfigure}
\usepackage{dcolumn}
\usepackage{bm}
\usepackage{amsfonts}
\usepackage{amssymb}
\usepackage{amsmath}
\usepackage{color}
\usepackage[colorlinks=true,breaklinks=true,linkcolor=blue,anchorcolor=red,citecolor=blue,urlcolor=blue]{hyperref}
\usepackage{longtable}
\usepackage{booktabs}
\usepackage{multirow}
\usepackage{changes}

\allowdisplaybreaks[3]

\def\bea{\begin{eqnarray}}
\def\eea{\end{eqnarray}}

\def\nn{\nonumber}
\def\Eq#1{Eq.~(\ref{#1})}

\def\Fig#1{Fig.~\ref{#1}}
\def\abs#1{\left|#1\right|}
\def\xk#1{\left(#1\right)}

\def\avg#1{\left\langle#1\right\rangle}

\newcommand{\s}{{\sigma}}

\newcommand{\de}{\delta}
\newcommand{\De}{\Delta}

\newcommand{\ga}{\gamma}
\newcommand{\Ga}{\Gamma}

\newcommand{\la}{\lambda}
\newcommand{\La}{\Lambda}

\renewcommand{\v}[1]{{\bf #1}}

\begin{document}

\title{Theory for Magnetic-Field-Driven 3D Metal-Insulator Transitions in the Quantum Limit}

\author{Peng-Lu Zhao}
\affiliation{Shenzhen Institute for Quantum Science and Engineering and Department of Physics, Southern University of Science and Technology (SUSTech), Shenzhen 518055, China}

\author{Hai-Zhou Lu}
\email{Corresponding author: luhz@sustech.edu.cn}
\affiliation{Shenzhen Institute for Quantum Science and Engineering and Department of Physics, Southern University of Science and Technology (SUSTech), Shenzhen 518055, China}
\affiliation{Shenzhen Key Laboratory of Quantum Science and Engineering, Shenzhen 518055, China}

\author{X. C. Xie}
\affiliation{International Center for Quantum Materials, School of Physics, Peking University, Beijing 100871, China}
\affiliation{CAS Center for Excellence in Topological Quantum Computation, University of Chinese Academy of Sciences, Beijing 100190, China}
\affiliation{Beijing Academy of Quantum Information Sciences, West Building 3, No. 10, Xibeiwang East Road, Haidian District, Beijing 100193, China}

\date{\today}

\begin{abstract}
Metal-insulator transitions driven by magnetic fields have been extensively studied in 2D, but a 3D theory is still lacking.
Motivated by recent experiments, we develop a scaling theory for the metal-insulator transitions in the strong-magnetic-field quantum limit
of a 3D system. By using a renormalization-group calculation to treat electron-electron interactions, electron-phonon interactions, and disorder on the same footing, we obtain the critical exponent that characterizes the scaling relations of the resistivity to temperature and magnetic field. By comparing the critical exponent with those in a recent experiment [\href{http://dx.doi.org/10.1038/s41586-019-1180-9}{F. Tang et al., Nature (London) 569, 537 (2019)}], we conclude that the insulating ground state was not only a charge-density wave driven by electron-phonon interactions but also coexisting with strong electron-electron interactions and backscattering disorder. We also propose a current-scaling experiment for further verification. Our theory will be helpful for exploring the emergent territory of 3D metal-insulator transitions under strong magnetic fields.
\end{abstract}

\maketitle

\begin{figure}[htbp]
\center
\includegraphics[width=0.46\textwidth]{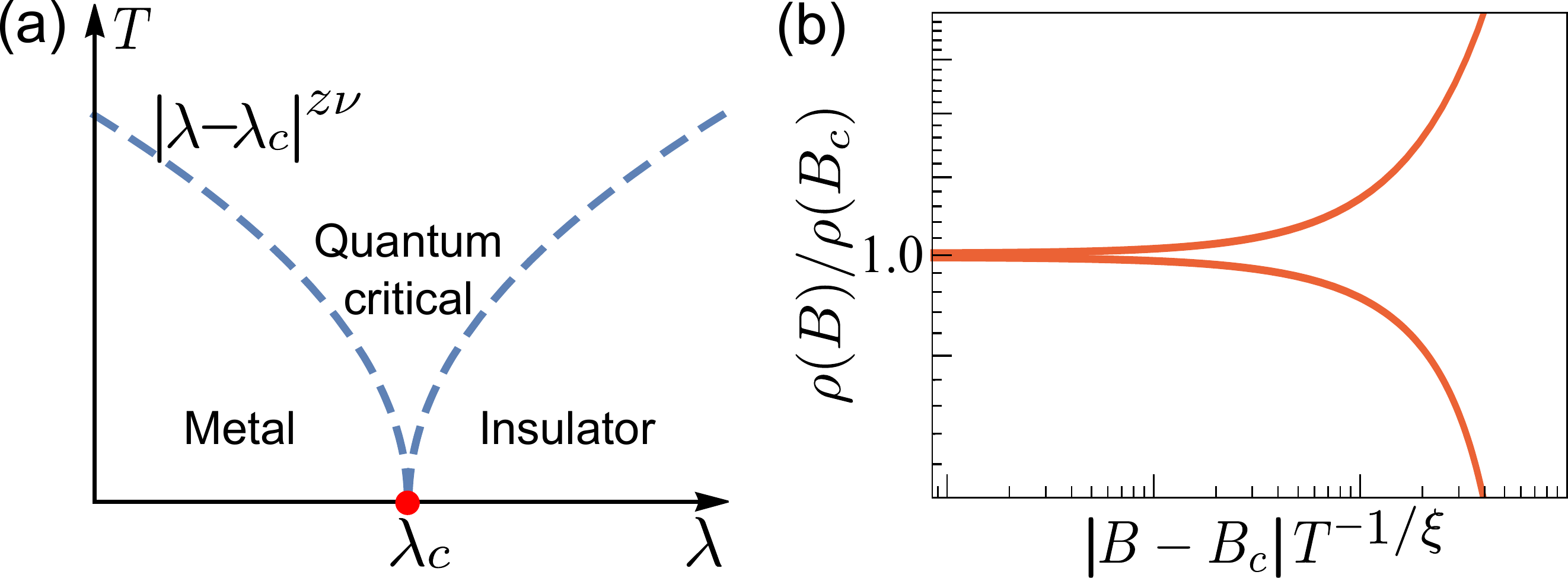}
\caption{(a) A quantum phase transition can be described by a nonthermal parameter $\la$ that converges to $\la_c$ at the quantum critical point. The boundaries (dashed curves) of the quantum critical region are described by $T \propto \abs{\la-\la_c}^{z\nu}$, where $z$ is the dynamical critical exponent and $\nu$ is the correlation length exponent. (b) The magnetic field $B$ can serve as $\lambda$, and $z$ and $\nu$ can be used to construct a measurable critical exponent $\xi$ that describes the resistivity as a scaling function of the magnetic field and temperature, e.g., $\rho(B)/\rho(B_c)=f(\abs{B-B_c}/T^{1/\xi})$, where $f(x)$ is a scaling function with $f(0)=1$. } \label{FigQCR}
\end{figure}

{\color{blue}\emph{Introduction.}}--Metal-insulator transition is a fascinating problem in condensed matter physics because of its rich mechanisms \cite{ImadaRMP,SondhiRMP,Kravchenko03}.
There have been extensive studies on 2D metal-insulator transitions in magnetic fields \cite{Newson86,Goldman86,Maliepaard89, Dai91,Kivelson92,Wangt94,Tomioka96,Popovif97,Xiexc98,AnJ01,Kempa02,Gorbar02,Kopelevich06,Geldart07,
ImadaRMP,SondhiRMP,Kravchenko03}, but a 3D theory is still lacking, despite that recent experiments show that magnetic fields can drive metal-insulator transitions in 3D systems \cite{Calder12,Ueda15,Tianzm16,Tangfd19,Wangp20} with characteristics of quantum phase transitions \cite{Vojta03, Sachdevb, Loehneysen07} [see \Fig{FigQCR}(a)], in particular, in the strong-field quantum limit of a topological insulator \cite{Tangfd19,Wangp20}. It is challenging to determine the insulating ground states in strong magnetic fields, because magnetic fields reduce the effective dimension, leading to stronger interactions and related instabilities, such as charge or spin density wave, Wigner crystal, Anderson localization, etc. \cite{Loehneysen07,Wigner34,Field86,Anderson58, Abrahams79,Evers08RMP,Giamarchib,Grunerb}.

\begin{figure}[htbp]
\center
\includegraphics[width=0.48\textwidth]{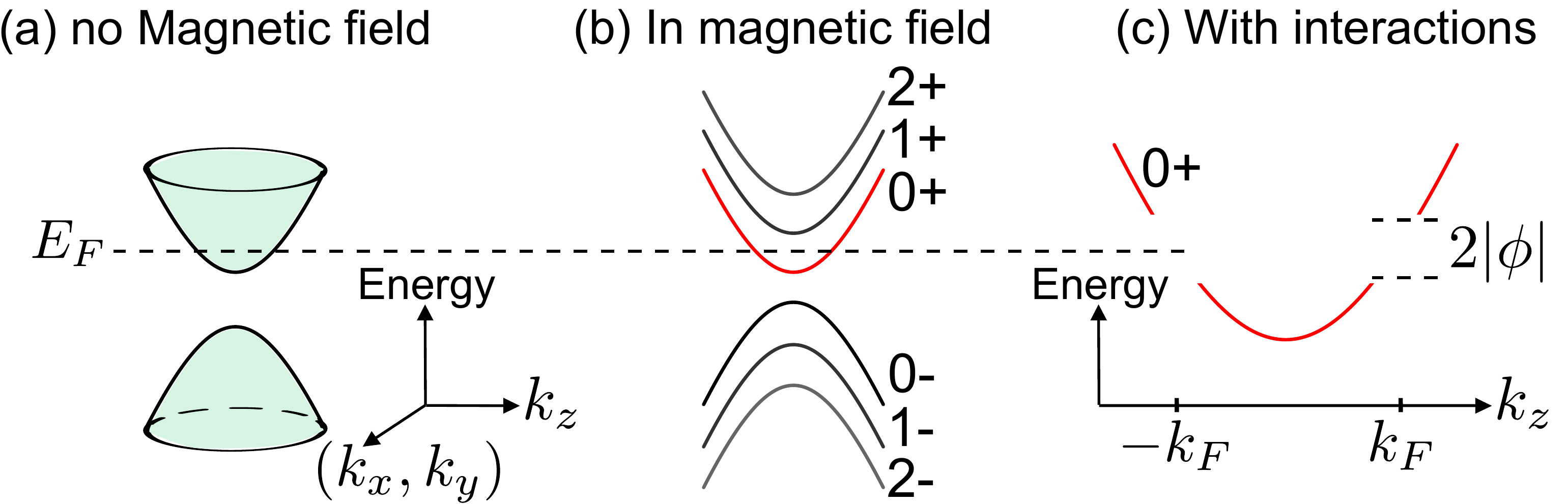}
\caption{A $z$-direction magnetic field splits the energy spectrum of a 3D insulator or metal (a) into 1D bands of Landau levels dispersing with the wave vector $k_z$ (b). The $n\pm$ stand for the $n$th Landau bands. In the quantum limit, the Fermi energy $E_F$ crosses the lowest Landau band $0+$ at $\pm k_F$, which is a metal phase and the starting point of this work. (c) Correlations between electrons near $-k_F$ and $k_F$ open a gap of size $2|\phi|$ and induce an insulating phase.}  \label{FigGaOp}
\end{figure}

In this Letter, we develop a scaling theory for the metal-insulator transition in the quantum limit of a 3D topological insulator under strong magnetic fields (see \Fig{FigGaOp}). Our theory agrees well with recent experiments~\cite{Tangfd19}.
The quantum phase transition is governed by universal relations described by critical exponents~\cite{Vojta03} [\Fig{FigQCR}(a)], corresponding to various instabilities and universal classes \cite{Wegner76,Abrahams79,McMillan81,Stafford93,Huckestein95,Dobrosavljevi97,Pelissetto02,Loehneysen07}.
With the help of a renormalization-group calculation, we find the dynamical critical exponents $z$ and correlation length exponents $\nu$ for the candidate instabilities. Based on them, we build the scaling relations of the resistivity to temperature and magnetic field, described by a critical exponent $\xi$. We find $\xi$ for five cases (Table \ref{tab:xi}).
By comparing with the experimentally measured $\xi=5.5$~\cite{Tangfd19}, we conclude that the insulating ground state is a charge-density wave not only driven by electron-phonon interactions but also requiring strong electron-electron interactions and backscattering disorder. We also propose a current-scaling experiment for further verification, by fitting the dynamical critical exponent and correlation length exponent. Our theory will be helpful for the emergent territory of 3D metal-insulator transitions under strong magnetic fields.

\begin{table}[htbp]
\caption{Comparison between experiments \cite{Tangfd19} and our theory on the critical exponent $\xi$ that describes the scaling relations of the resistivity to temperature and magnetic field, for different dominant and coexisting interactions and disorder. Theoretically, $\xi$ is a product of the  dynamical  critical  exponents $z$ and correlation length exponents $\nu$.
In the experiments, the metal-insulator transition happens at a critical magnetic field of $B_c= 6.71$ T, where an incommensurate charge-density wave dominates the ground state according to our theory \cite{Supp},
different from the previous work \cite{Qin20prl} for $B\in[1.7,2.1]$ T, where the Hall resistivity plateau indicates a commensurate charge-density wave as the ground state. \label{tab:xi}}
\begin{ruledtabular}
\begin{tabular}{lllllc}
Insulating & Dominant  & Coexisting & $z$ & $\nu$ &  $\xi$  \\
phases &    &   &  &   &   \vspace{0.2cm}\\
\hline
Charge- & Electron- & Electron- & 1&1.5&1.5
\\
density  &phonon &  electron & &  &
\\
wave
&  &  & & &
\vspace{0.2cm}\\
Anderson & Forward-  & Electron-phonon  & 1& 2& 2
\\
insulator &scattering   & Electron- &  &  &
\\
  & disorder &electron  &
\vspace{0.2cm}\\
Charge- & Electron  & Backscattering& 1.5 & 3 & 4.5
\\
density& -phonon  & disorder &  &  &
\\
 wave & &Weak electron & & &
\\
& &-electron & & &
\vspace{0.2cm}\\
Wigner & Electron & Backscattering &  2 & 1& 2
\\
crystal & -electron & disorder&  & &
\\
&   & Electron- & & &
\\
&   & electron  & & &
\vspace{0.2cm}\\
Charge- & Electron & Backscattering & 2 & 3 & 6
\\
density & -phonon  & disorder &  &  &
\\
wave &  &  Strong electron  & & &
\\
 &  &  -electron & & &
\end{tabular}
\\
Experiment
\begin{tabular}{cccccccc}
Sample &1 &2& 3  &4 &Mean value &Mean value\\
& &&  & & (all samples) &(2,3,4)\\ \hline
$\xi$  &3.95 &6.06&5.78& 6.25&5.5 $\pm$ 1.1 & 6.0 $\pm$ 0.1
\tabularnewline
\end{tabular}\label{tab:samples}
\end{ruledtabular}
\end{table}

{\color{blue}\emph{Model for 3D systems in strong fields.}}--3D insulators and metals can be generically described by a Dirac model \cite{Shen17book}
\bea
\mathcal{H}=m(\mathbf{k})\tau_{z}\sigma_{0}+v_{x}k_{x}\tau_{x}\sigma_{z}+v_{y}k_{y}\tau_{y}\sigma_{0}+v_{z}k_{z}\tau_{x}\sigma_{x},
\eea
where $\mathbf{k}=(k_x,k_y,k_z)$ is the wave vector, $\tau$ and $\sigma$ are Pauli matrices for pseudo and real spin, respectively, $v_{x,y,z}$ are parameters for the ``Fermi velocities," and $m(\mathbf{k})$ stands for the ``Dirac mass" \cite{Chenry15, JiangY17}. We will focus on the quantum limit in strong fields, where a $z$-direction magnetic field splits the energy spectrum into a series of 1D bands of Landau levels dispersing with $k_z$ and only the lowest Landau band $0+$ is occupied (Fig.~\ref{FigGaOp}).
An effective free Hamiltonian for the lowest Landau band can be found by linearizing the dispersion
near the two Fermi wave vectors $\pm k_{F}{\bf e}_{z}$ [\Fig{FigGaOp}(b)] as
$H_{0}=\sum_{\v k}\psi^{\dagger}\xk{\bf k} v_{F} k_{z}\s_z \psi\xk{\bf k}$ \cite{Supp},
where $\psi\xk{\bf k}=[\psi_{-}\xk{\bf k},\psi_{+}\xk{\bf k}]^{\text{T}}$, $\psi_{\pm}^{\dagger}\xk{\bf k}$ and $\psi_{\pm}\xk{\bf k}$ are the creation and annihilation operators, respectively, near $\mp k_{F}$,
and $\bf k$ is measured from $\pm k_{F}{\bf e}_{z}$.

{\color{blue}\emph{Interactions and disorder.}}--Electron-electron or electron-phonon interactions can open a charge-density-wave gap near $\pm k_F$ [see \Fig{FigGaOp}(c)]. The charge-density wave may induce the Wigner crystal \cite{Wigner34,Field86}. Also, disorder could induce the Anderson localization \cite{Anderson58, Abrahams79,Evers08RMP}.  We study them on the same footing by using the effective action
\bea
S&=&\int d^{3} x  d \tau (\mathcal{L}_{e}+\mathcal{L}_{p}+\mathcal{L}_{b})+ \int d^{3} x  d \tau  d \tau' \mathcal{L}_{d} ,
\label{Eqactint} \\
\mathcal{L}_{e}&=&u \psi_{+}^{\dagger}\psi_{+} \psi_{-}^{\dagger}\psi_{-},
\ \
\mathcal{L}_{p}= g (\phi \psi_{+}^{\dagger} \psi_{-}+\phi^{\ast} \psi_{-}^{\dagger} \psi_{+} ),
\nonumber\\
\mathcal{L}_{b}&=&   \left|\partial_{\tau} \phi \right|^{2}+ v_b^2\left|\bm{\partial} \phi\right|^{2}
+r|\phi|^{2},
\nonumber\\
\mathcal{L}_{d}&=&-\sum_{i=f,b}(\De_i/2) \xk{\psi^{\dagger}\Ga_i\psi}_{\tau} \xk{\psi^{\dagger}\Ga_i\psi}_{\tau'} \nn.
\eea
The Lagrangian $\mathcal{L}_{e}$ describes electron-electron interactions that induce the $2k_F$ instability \cite{Jiansk20}. $\mathcal{L}_{p}$ represents the coupling between charge-density-wave order parameter $\phi$ and electrons \cite{Grunerb,Bruusb,Qin20prl}. The order parameter is defined as $\phi=(\alpha_{2k_F}/ V)(\langle b_{-2 k_{F} \mathbf{e}_{z}}^{\dagger} \rangle+ \langle b_{2 k_{F} \mathbf{e}_{z}} \rangle )$,
where $b_{\mathbf{q}}^{\dagger}$ and $b_{\mathbf{q}}$ are the creation and annihilation operators, respectively, for the phonons with momentum $\mathbf{q}$ and $\alpha_{q}$ measures the electron-phonon coupling strength \cite{Bruusb,Qin20prl}.
$\mathcal{L}_{b}$ describes the order parameter dynamics \cite{Sachdevb}
with $r=0$ at the quantum critical point.
$\mathcal{L}_{d}$ is the Lagrangian of disorder after being ensemble averaged by means of the replica method~\cite{Wegner79, Efetov80, Lee85}. The Hamiltonian of disorder takes the form $\mathcal{H}_{\text{dis} }=U_i(x)\psi^{\dagger} \Ga_i \psi$, where $U_i(x)$ is the impurity potential of a Gaussian white-noise distribution as $\avg{U_i}=0$ and $\avg{U_i(\v x)U_j\xk{\v x'}} =\De_i\de_{ij} \de(\v x-\v x')$.
For the forward-scattering disorder, $\Ga_i=\s_0$, which couples electrons near $k_F$ only to those near $k_F$ or $-k_F$ only to those near $-k_F$ \cite{Giamarchi88,Giamarchib}, which may induce the Anderson localization \cite{Anderson58, Abrahams79,Evers08RMP}.
For the backscattering disorder, $\Ga_i=\s_x +\s_y$, which couples electrons near $k_F$ to those near $-k_F$ or $-k_F$ to those near $k_F$ \cite{Giamarchi88,Giamarchib}.
We study these two types of disorder separately.

{\color{blue}\emph{Renormalization-group equations.}}--The renormalization group \cite{Fisher74,Wilson83,MacKinnon94,Stanley99} is an approach to determine the instabilities and corresponding critical exponents. We perform a Wilsonian momentum-shell renormalization-group analysis \cite{Wilson74,Abrikosovb,Peskinb,Mahanb,Shankar94} for the model described by \Eq{Eqactint}. The momentum shell is defined as $e^{-\ell}\Lambda <|\v k_z| < \Lambda$, where $\ell$ is the running scale parameter.
The renormalization-group flow equations are found as \cite{Supp}
\bea
d v_F/d\ell&=&[z-1+ 2\beta\ga^3/ \xk{\ga+1}^2-a_{1 i} \De_i]v_F,
\nonumber\\
d \De_f/d\ell&=&2\De_f^2+\De_f - 4\beta \ga^3\De_f/\xk{\ga+1}^2,
\nonumber\\
d \De_b/d\ell&=&-4\De_b^2+\xk{ 1+2u}\De_b-\frac{2\beta \ga^2\xk{2\ga+1}\De_b}{\xk{\ga+1}^2},
\nonumber\\
d \beta/d\ell&=&- 2\beta^2 \ga^2/\xk{\ga+1}^2+
\xk{2+2u-2a_{2 i}\De_i }\beta,
\nonumber\\
d u/d\ell&=&
2u^2- 2\beta\ga^2 u/(\ga+1)-a_{3 i}\De_i u,\label{EqRGu}
\eea
where $a_{mi}$ is a $3\times 2$ constant matrix with elements $a_{mi}=(m,2)$, and $i=f,b$ representing the forward- and backward-scattering disorder, respectively. $z$ is the dynamic exponent, which can be found by fixing $v_F$. We have redefined the dimensionless coupling constants as
\begin{eqnarray}
 \textrm{electron-phonon:}\quad && g^2/4\pi^2\ell_B^2 v_b^3\La^2 \rightarrow \beta, \nn \\
 \textrm{electron-electron:}\quad && u/4\pi^2v_F\ell_B^2 \rightarrow u,\nn\\
 \textrm{disorder:}\quad && \De_i/2\pi^2 v_F^2 \ell_B^2 \La\rightarrow \De_i, \label{Eqeffectcoupling}
\end{eqnarray}
where $\ell_B=\sqrt{\hslash/(eB)}$ is the magnetic length. Different from nonmagnetic field theories, the
integrals in the plain give the Landau degeneracy of
the magnetic field \cite{Supp}.
The renormalization-group equations highly depend on the parameter $\gamma\equiv|v_b/v_F|$, the ratio of the phonon speed to the Fermi velocity, which does not flow with $\ell$, and $\ga \approx 1.7\times 10^{-3}$ from the model parameters \cite{Supp,Zhuj18,Abrikosov98,Luhz15}. Later, our conclusions hold for a range of small $\gamma$.

\begin{figure}[htbp]
\center
\includegraphics[width=0.48\textwidth]{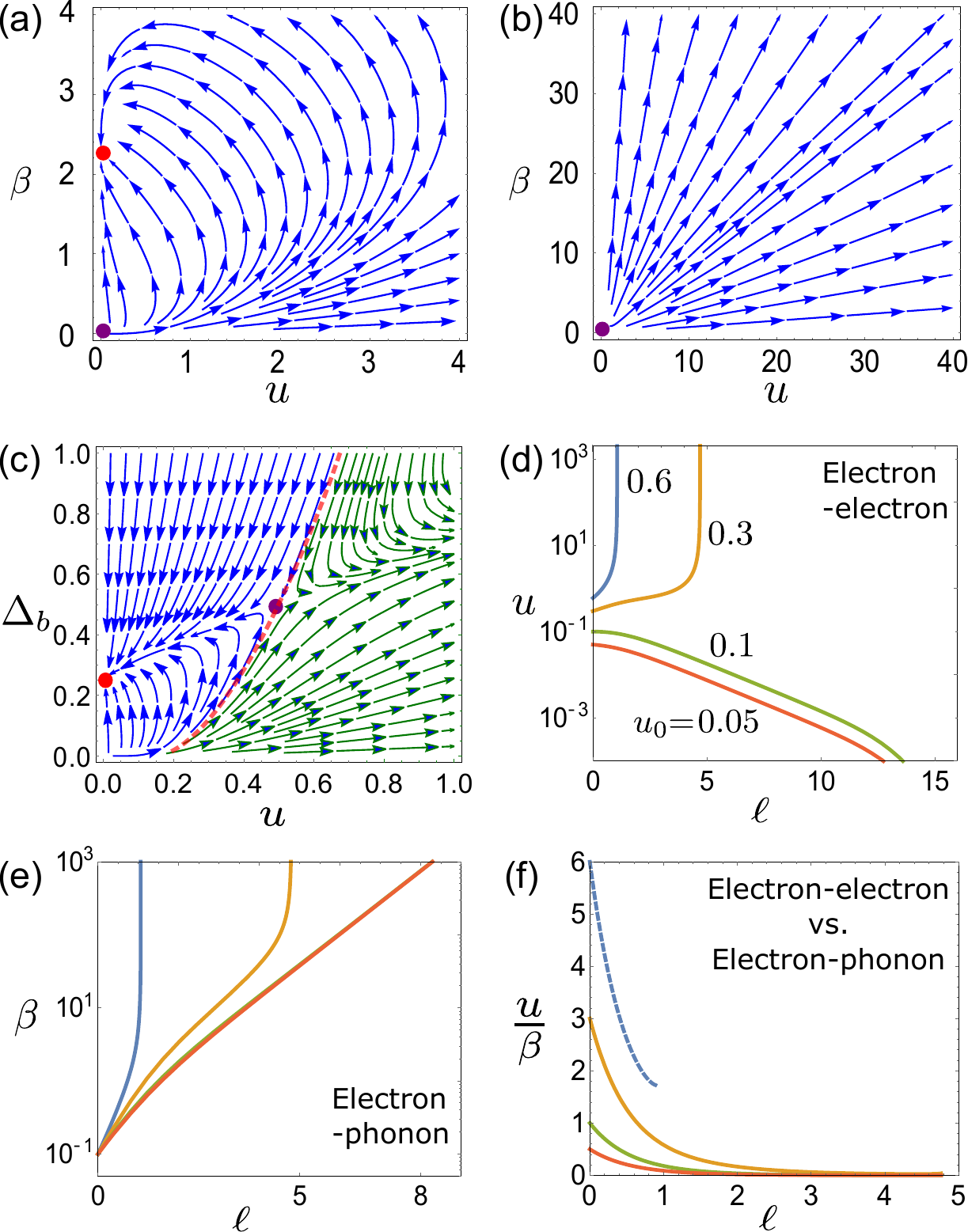}
\caption{The renormalization-group flow in the $u-\beta$ plane for (a) $\ga=2$ and (b) $\ga=0.002$.
$u$ and $\beta$ measure the effective electron-electron and electron-phonon interactions, respectively.
$\ga=|v_b/v_F|$ is the ratio of the phonon speed to Fermi velocity.
There are two fixed points at (i)$_c$ $(u_\ast,\beta_\ast)=(0,0)$ and (ii)$_c$ $[0,\xk{1+1/\ga}^2]$.
(c) The renormalization-group flow in the $u$-$\De_b$ plane, in the absence of electron-phonon coupling. In this case, there are two non-Gaussian fixed points at $\xk{u_{\ast},\De_b^{\ast}}=\xk{0,1/4}$ and $\xk{1/2,1/2}$. The red curve distinguishes a disordered metal with zero electron-electron coupling on the left and a Wigner crystal on the right \cite{Supp}. (d)-(f) Running of the electron-electron coupling $u$, electron-phonon coupling $\beta$, and $u/\beta$ with $\ell$,
for different initial values of $u$. The initial values of both $\De_b$ and $\beta$ are 0.1. (d)--(f) share the same legends.
The green and red lines overlap in (e).} \label{Figbd}
\end{figure}

{\color{blue}\emph{Charge-density wave without disorder?}}--For a system with only electron-electron and electron-phonon interactions, there are two fixed points [\Fig{Figbd}(a)] at (i)$_c$ $(u_{\ast},\beta_{\ast})=(0,0)$ and (ii)$_c$  $[0,\xk{1+1/\ga}^2]$.
If the phonon-Fermi velocity ratio is not extremely small  (e.g., $\ga >1$), the attractive fixed point (ii)$_c$  \cite{Supp} has a small value for electron-phonon interactions. Neither electron-electron nor electron-phonon interactions could induce the charge-density wave. This conclusion explicitly conflicts with the Peierls phase transition \cite{Grunerb,Kohn59}. Because of an extremely small $\ga$, the attractive fixed point (ii)$_c$  behaves like an infinitely large electron-phonon coupling point. Before reaching the attractive point (ii)$_c$, the electron-phonon coupling becomes strong enough to induce the Peierls phase transition, which opens a charge-density-wave gap. Therefore, an extremely small $\gamma$ is crucial to induce the Peierls phase transition. In this case, electron-phonon interactions can be viewed as a relevant perturbation starting from the Gaussian fixed point (i)$_c$ [see \Fig{Figbd}(b)]. The linearized renormalization-group equations around the fixed point (i)$_c$ show that the electron-phonon coupling increases with $\ell$ as $e^{2\ell}$. Therefore, the correlation length diverges with an exponent $\nu_{\beta}=1/2$ and the dynamical critical exponent $z=1$ \cite{Supp,Goswami11L}. The general scaling relation of the resistivity to temperature and magnetic field is given by \cite{SondhiRMP,Fisher90,ImadaRMP}
\bea
\rho(B, T)=\rho (B_c)f[\abs{B-B_{c}} / T^{1/z\nu_B }],
\eea
where $f(x)$ is a scaling function with $f(0)=1$.
To obtain the general scaling with respect to the magnetic field, we have defined $\nu_B$, which describes how the correlation length $\zeta$ diverges with the magnetic field, following $\zeta \sim |B-B_c|^{-\nu_{B}}$. On the other hand, the correlation length exponent obtained by our renormalization-group analysis describes that the correlation length $\zeta$ diverges with the coupling strength as $\zeta \sim |\beta-\beta_c|^{-\nu_{\beta}}$.
Our scaling analysis
based on \Eq{Eqeffectcoupling}
yields $\nu_{B}=3\nu_{\beta}$ \cite{Supp}, and the critical exponent takes the form
\bea
\xi=3z\nu_{\beta}=1.5.
\eea
This value is far less than the experimental value $\xi=5.5$ ~\cite{Tangfd19}, so without disorder, the coexistence of electron-electron and electron-phonon interactions cannot induce the metal-insulator transitions in the experiments \cite{Tangfd19}.

{\color{blue}\emph{Anderson localization induced by forward-scattering disorder?}}--When forward-scattering disorder exists, Anderson localization happens first with $z=1, \nu_{B}=2\nu_{\De}=2$, and $\xi=2$ (see details in Sec. SIIIE in \cite{Supp}), which is far less than the experimental value $\xi=5.5$~\cite{Tangfd19}.

{\color{blue}\emph{Charge-density wave with backscattering disorder.}}--When backscattering disorder exists, the fixed points are (i)$_b$ $\xk{u_{\ast},\beta_{\ast},\De_b^{\ast}}=\xk{0,0,0}$; (ii)$_b$ $[0,\xk{1+1/\ga}^2,0]$; (iii)$_b$ $\xk{0,0,1/4}$; and (iv)$_b$ $\xk{1/2,0,1/2}$, belonging to four different universality classes because they have different critical exponents. Fixed point (i)$_b$ is unstable for electron-phonon interactions and disorder. Fixed point (ii)$_b$ is attractive for all three couplings and represents the charge-density wave with an irrelevant electron-electron coupling and backscattering disorder \cite{Supp}. Without electron-phonon interactions, fixed point (iii)$_b$ is stable and fixed point (iv)$_b$ is a critical point for a Wigner crystal \cite{Supp}. As shown in \Fig{Figbd}(c), there exists a critical line in the $u$-$\De_b$ plane; on the right, $u$ flows to infinity, indicating a charge-density wave driven by electron-electron interactions, i.e., a Wigner crystal, while on the left, the system flows to a finite disorder fixed point with a stable zero-valued electron-electron coupling.
When including electron-phonon interactions, as shown in \Fig{Figbd}(d), the behavior of $u$ does not change qualitatively. Figure \ref{Figbd}(e) shows that the electron-phonon coupling $\beta$ flows to infinity with increasing $\ell$. Near fixed point (iii)$_b$, $\beta$ is the only unstable coupling. Fixed point (iii)$_b$ becomes the critical fixed point for a charge-density wave induced by electron-phonon interactions in the presence of a finite backscattering disorder and irrelevant electron-electron coupling.
The divergent $\beta$ gives $\nu_{\beta}=1$ \cite{Supp}.
The dynamical critical exponent for this universality class is found as
$z=1+2\De_b^{\ast}=1.5$. We thus get a critical exponent for the resistivity scaling
\bea
\xi=3z\nu_{\beta}=4.5,
\eea
which is close to the experimental value $\xi=5.5$ \cite{Tangfd19}.

Near fixed point (iv)$_b$, both $\beta$ and $u$ are unstable and increase unboundedly when the initial value of $u$ is large enough. Fixed point (iv)$_b$ is a multicritical point for the Wigner crystal or Peierls phase transition. Which phase transition occurs first depends on the relative strength of $\beta$ and $u$. The solid lines in \Fig{Figbd}(e) show that the ratio of $u$ to $\beta$ decreases rapidly and finally vanishes when its initial value is not large enough, indicating that the Peierls phase transition happens first. The dotted line in \Fig{Figbd}(f) shows that $u$ is always larger than $\beta$, when its initial value is large enough, which means that the Wigner crystal occurs first. Despite that these two phase transitions both produce charge-density waves, the gap sizes and critical behaviors are different \cite{Grunerb,Giamarchib}.
For the Peierls phase transition, the correlation length exponent is $\nu_{\beta}=1$, and the dynamical critical exponent is
$z=1+2\De_b^{\ast}=2$, which gives a critical exponent for the resistivity scaling:
\begin{eqnarray}\label{xi-6}
\xi=3z\nu_{\beta}=6.
\end{eqnarray}
For the Wigner crystal, we obtain $\nu_u=1$ and $z=2$, which lead to $\xi=2$ \cite{Supp}.

\textcolor[rgb]{0.00,0.00,1.00}{\emph{Conclusion and experimental verification.}}--So far, $\xi=6$ is closest to the experimental result $\xi=5.5$ \cite{Tangfd19}.
Therefore, we conclude that the magnetic-field-induced metal-insulator transition in the experiments is likely a charge-density wave induced by electron-phonon interactions, in the presence of strong electron-electron interactions and backscattering disorder (only for samples 2--4 in Ref. \cite{Tangfd19}; see discussion below). Its quantum critical behavior is described by the universal class of fixed point (iv)$_b$ with a correlation length exponent $\nu_B=3$ and dynamical critical exponent $z=2$.

\begin{figure}[htbp]
\center
\includegraphics[width=0.45\textwidth]{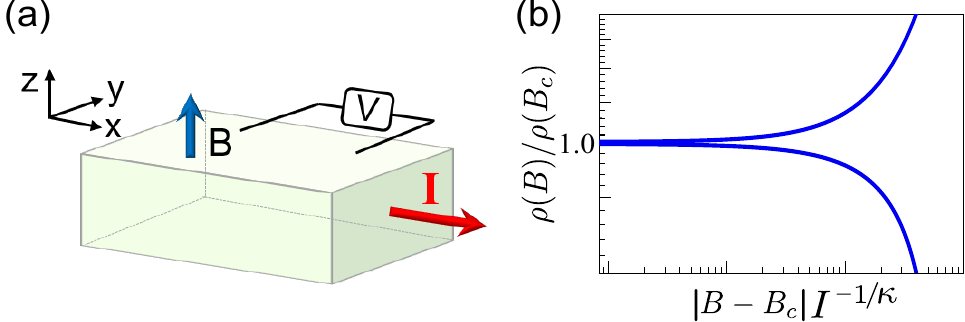}
\caption{(a) The experimental setup
for the current-scaling measurement.
The longitudinal resistivity $\rho\equiv V/I$ along the $x$ direction as a function of the magnetic field $B$ is measured at different measurement currents $I$ to construct a scaling function $\rho(B)/\rho(B_c)=\mathcal{F}(\abs{B-B_c}/I^{1/\kappa})$ in (b), where $\mathcal{F}(x)$ is a scaling function with $\mathcal{F}(0)=1$.
The linear fitting of the $\log(d\rho/dB|_{B=B_c})$ vs. $--\log(I)$ yields
$1/\kappa$.
With the fitted $\xi$ in Fig. \ref{FigQCR} and $\kappa$ here, the correlation length exponent $\nu_B=\kappa-\xi$ and the dynamical critical exponent $z=\xi/(\kappa-\xi)$ can be found, to provide more experimental connections for our theory to identify the insulating ground state in the experiments.
} \label{Exptest}
\end{figure}

Our conclusion can be verified by performing the current-scaling measurement  \cite{Polyakov93B,WeiHP94B,Kravchenko96L,Panw97B,Salehi19AM}, which along with the temperature scaling results in Ref. \cite{Tangfd19} could give rise to the experimental values of $z$ and $\nu$. The experimental setup is shown in \Fig{Exptest}(a). One needs to measure the longitudinal resistivity as a function of the magnetic field at different measurement currents $I$. Rearranging the measured magnetoresistivity $\rho(B)/\rho(B_c)$ near the critical field $B_c$ as a function of $\abs{B-B_c}I^{-1/\kappa}$ could generate a current-scaling plot as shown in \Fig{Exptest}(b), which is similar to that in Fig. \ref{FigQCR} (b). The scaling relation of the resistivity to the current takes the form \cite{SondhiRMP}
\bea
\rho(B, T)=\rho (B_c)\mathcal{F}[\abs{B-B_{c}} / I^{1/(z+1)\nu_B }],
\eea
where $\mathcal{F}(x)$ is a scaling function with $\mathcal{F}(0)=1$.
With the fitted $\xi$ in Ref. \cite{Tangfd19} and $\kappa$ in this current-scaling measurement, the correlation length exponent $\nu_B$ and dynamical critical exponent $z$ can be found as
\begin{eqnarray}
\nu_B=\kappa-\xi,\ \ \
z=\xi/(\kappa-\xi),
\end{eqnarray}
respectively.
Previously, this method has been widely employed in 2D quantum phase transitions \cite{WeiHP94B,Kravchenko96L,Panw97B,Salehi19AM}, and it could provide an experimental way of distinguishing our theoretical exponents in Table~\ref{tab:xi}.

Also, the charge-density wave could be probed in X-ray diffraction spectrum \cite{Wangs15RSI,Narumi06JPCS,Pototsching97}, though the combination of the high magnetic field and X-ray facilities is challenging.

\textcolor[rgb]{0.00,0.00,1.00}{\emph{Discussion.}}--Our theory provides a deeper understanding to the data in Ref.~\cite{Tangfd19}, where the value $\xi=5.5$ is obtained by averaging four samples, as shown in Table~\ref{tab:xi}.
However, the value of $\xi$ for sample 1 is closer to the value $\xi=4.5$ of the universal class described by fixed point (iii)$_b$. By contrast, $\xi$ of samples 2--4 belongs to fixed point (iv)$_b$.
The physical difference is that electron-electron interactions in sample 1 can be seen as zero valued when the metal-insulator transition happens, whereas they are finite and strong in the other three samples. As shown in Table~\ref{tab:xi}, the mean value of samples 2--4 gives rise to $\xi=6.0$, which remarkably agrees with our theoretical value in \Eq{xi-6}. Despite that we take the experiments in Ref.~\cite{Tangfd19} as a concrete sample to compare with, our theory can be applied to other 3D metal-insulator transitions under strong magnetic fields. Recently, the field-driven metal-insulator transition in $\beta-\text{Bi}_4\text{I}_4$ \cite{Wangpp21B} is found to have $\xi=6.5$ within the experimental error, showing the wide application potential of our theory.
Nevertheless, 3D metal-insulator transitions of spin-correlated systems driven by magnetic fields \cite{Calder12,Ueda15,Tianzm16} are beyond the scope of our theory and will be a challenging topic in the future.

\begin{acknowledgments}
We thank helpful discussions with Liyuan Zhang and Tingyu Gao. This work was supported by the National Natural Science Foundation of China (11925402 and 12047531), the National Basic Research Program of China (2015CB921102), the Strategic Priority Research Program of Chinese Academy of Sciences (XDB28000000), Guangdong province (2020KCXTD001 and 2016ZT06D348), Shenzhen High-level Special Fund (G02206304 and G02206404), and the Science, Technology and Innovation Commission of Shenzhen Municipality (ZDSYS20170303165926217, JCYJ20170412152620376, and KYTDPT20181011104202253). The numerical calculations were supported by Center for Computational Science and Engineering of SUSTech.
\end{acknowledgments}

\bibliography{CDWMIT,refs-transport}

\end{document}